\newcommand{\psfile}[3][]{ 
  \begin{center}
    \setlength{\epsfxsize}{#3\linewidth}\leavevmode
    \def\noOpt{}\def\testit{#1}\ifx\testit\noOpt%
      \epsfbox{#2}%
    \else%
      \epsfbox[#1]{#2}%
    \fi
  \end{center} 
}

\newcommand{\goto}{\rightarrow}
\newcommand{\calB}{\mbox{${\cal B}$}}
\newcommand{\calL}{\mbox{${\cal L}$}}
\newcommand{\calP}{\mbox{${\cal P}$}}
\newcommand{\vbeta}{\mbox{$\vec\beta$}}
\newcommand{\vgamma}{\mbox{$\vec\gamma$}}
\newcommand{\etaK}{\mbox{$B\goto\eta K$}}
\newcommand{\etapK}{\mbox{$B\goto\eta^\prime K$}}
\newcommand{\etapKp}{\mbox{$B^+\goto\eta^\prime K^+$}}
\newcommand{\BetapKp}{\mbox{$\calB(B^+\goto\eta^\prime K^+)$}}
\newcommand{\RetapKp}{\mbox{$(6.5^{+1.5}_{-1.4}\pm0.9)\times10^{-5}$}}
\newcommand{\etapKz}{\mbox{$B^0\goto\eta^\prime K^0$}}
\newcommand{\BetapKz}{\mbox{$\calB(B^0\goto\eta^\prime K^0)$}}
\newcommand{\RetapKz}{\mbox{$(4.7^{+2.7}_{-2.0}\pm0.9)\times10^{-5}$}}
\newcommand{\DE}{\mbox{$\Delta E$}}
\newcommand{\mb}{\mbox{$M$}}
\newcommand{\xf}{\mbox{${\cal F}$}}
\newcommand{\hel}{\mbox{${\cal H}$}}
\newcommand{\GeVc}{\mbox{${\rm GeV}/c$}}

\newcommand{\MeVc}{\mbox{${\rm MeV}/c$}}
\newcommand{\MeVcsq}{\mbox{${\rm MeV}/c^2$}}

\def\sgline{\noalign{\vskip 0.15truecm\hrule\vskip 0.15truecm}}
\newcommand{\piz}{\mbox{$\pi^0$}}
\newcommand{\etapr}{{\eta^{\prime}}}

\newcommand{\etaprkp}{\mbox{$\etapr K^+$}}
\newcommand{\etaprkpd}{\mbox{$\etapr_{\eta\pi\pi}K^+$}}
\newcommand{\etaprkprg}{\mbox{$\etapr_{\rho\gamma}K^+$}}
\newcommand{\etaprkpfv}{\mbox{$\etapr_{5\pi}K^+$}}
\newcommand{\etaprkz}{\mbox{$\etapr K^0$}}
\newcommand{\etaprkzd}{\mbox{$\etapr_{\eta\pi\pi} K^0$}}
\newcommand{\etaprkzrg}{\mbox{$\etapr_{\rho\gamma} K^0$}}
\newcommand{\etaprpi}{\mbox{$\etapr\pi^+$}}
\newcommand{\etaprpid}{\mbox{$\etapr_{\eta\pi\pi}\pi^+$}}
\newcommand{\etaprpirg}{\mbox{$\etapr_{\rho\gamma}\pi^+$}}
\newcommand{\etaprpifv}{\mbox{$\etapr_{5\pi}\pi^+$}}

\newcommand{\etaprpiz}{\mbox{$\etapr\piz$}}
\newcommand{\etaprpizepp}{\mbox{$\etapr_{\eta\pi\pi}\piz$}}
\newcommand{\etaprpizrg}{\mbox{$\etapr_{\rho\gamma}\piz$}}
\newcommand{\etaprkstz}{\mbox{$\etapr K^{*0}$}}
\newcommand{\etaprkstzd}{\mbox{$\etapr_{\eta\pi\pi} K^{*0}$}}
\newcommand{\etaprkstp}{\mbox{$\etapr K^{*+}$}}
\newcommand{\etaprkstpd}{\mbox{$\etapr_{\eta\pi\pi} K^{*+}_{K^+\piz}$}}
\newcommand{\etaprkstpkz}{\mbox{$\etapr_{\eta\pi\pi} K^{*+}_{K^0\pi^+}$}}
\newcommand{\etaprrhoz}{\mbox{$\etapr\rho^0$}}
\newcommand{\etaprrhozd}{\mbox{$\etapr_{\eta\pi\pi}\rho^0$}}
\newcommand{\etaprrhop}{\mbox{$\etapr\rho^+$}}
\newcommand{\etaprrhopd}{\mbox{$\etapr_{\eta\pi\pi}\rho^+$}}
\newcommand{\etapreta}{\mbox{$\etapr\eta$}}
\newcommand{\etapretagg}{\mbox{$\etapr_{\eta\pi\pi}\eta_{\gaga}$}}
\newcommand{\etapretathrp}{\mbox{$\etapr_{\eta\pi\pi}\eta_{3\pi}$}}
\newcommand{\etapretarg}{\mbox{$\etapr_{\rho\gamma}\eta_{\gaga}$}}
\newcommand{\etapretargtp}{\mbox{$\etapr_{\rho\gamma}\eta_{3\pi}$}}
\newcommand{\etapretapr}{\mbox{$\etapr\etapr$}}
\newcommand{\etapretaprd}{\mbox{$\etapr_{\eta\pi\pi}\etapr_{\eta\pi\pi}$}}
\newcommand{\etapretaprrg}{\mbox{$\etapr_{\eta\pi\pi}\etapr_{\rho\gamma}$}}

\newcommand{\etak}{\mbox{$\eta K^+$}}
\newcommand{\etakgg}{\mbox{$\eta_{\gaga} K^+$}}
\newcommand{\etakthrp}{\mbox{$\eta_{3\pi} K^+$}}
\newcommand{\etapi}{\mbox{$\eta\pi^+$}}
\newcommand{\etapigg}{\mbox{$\eta_{\gaga}\pi^+$}}
\newcommand{\etapithrp}{\mbox{$\eta_{3\pi}\pi^+$}}
\newcommand{\etapiz}{\mbox{$\eta\piz$}}
\newcommand{\etapizgg}{\mbox{$\eta_{\gaga}\piz$}}
\newcommand{\etapizthrp}{\mbox{$\eta_{3\pi}\piz$}}
\newcommand{\etakz}{\mbox{$\eta K^0$}}
\newcommand{\etakzgg}{\mbox{$\eta_{\gaga} K^0$}}
\newcommand{\etakzthrp}{\mbox{$\eta_{3\pi} K^0$}}
\newcommand{\etaeta}{\mbox{$\eta\eta$}}
\newcommand{\etaetagg}{\mbox{$\eta_{\gaga}\eta_{\gaga}$}}
\newcommand{\etaetathrp}{\mbox{$\eta_{\gaga}\eta_{3\pi}$}}
\newcommand{\etaetasixp}{\mbox{$\eta_{3\pi}\eta_{3\pi}$}}
\newcommand{\etakstz}{\mbox{$\eta K^{*0}$}}
\newcommand{\etakstzgg}{\mbox{$\eta_{\gaga} K^{*0}$}}
\newcommand{\etakstzthrp}{\mbox{$\eta_{3\pi} K^{*0}$}}
\newcommand{\etakstp}{\mbox{$\eta K^{*+}$}}
\newcommand{\etakstpgg}{\mbox{$\eta_{\gaga} K^{*+}_{K^+\piz}$}}
\newcommand{\etakstpthrp}{\mbox{$\eta_{3\pi} K^{*+}_{K^+\piz}$}}
\newcommand{\etakstpggkz}{\mbox{$\eta_{\gaga} K^{*+}_{K^0\pi^+}$}}
\newcommand{\etakstpthrpkz}{\mbox{$\eta_{3\pi} K^{*+}_{K^0\pi^+}$}}
\newcommand{\etarhoz}{\mbox{$\eta \rho^0$}}
\newcommand{\etarhozgg}{\mbox{$\eta_{\gaga} \rho^0$}}
\newcommand{\etarhozthrp}{\mbox{$\eta_{3\pi} \rho^0$}}
\newcommand{\etarhop}{\mbox{$\eta \rho^+$}}
\newcommand{\etarhopgg}{\mbox{$\eta_{\gaga} \rho^+$}}
\newcommand{\etarhopthrp}{\mbox{$\eta_{3\pi} \rho^+$}}

\newcommand{\gaga}{{\gamma\gamma}}

\documentstyle[aps,prl,preprint,floats,epsfig]{revtex}

\begin{document}

\preprint{\tighten\vbox{\hbox{\hfil CLNS 97/1536}
                        \hbox{\hfil CLEO 97-31}
}}

\title{
Two-body \boldmath{$B$} Meson Decays to \boldmath{$\eta$} and
\boldmath{$\eta^\prime$}--Observation of \boldmath{$B\goto\eta^\prime K$}
}  

\author{CLEO Collaboration}
\date{January 12, 1998}

\maketitle
\tighten

\begin{abstract} 
In a sample of 6.6 million produced $B$ mesons we have observed decays
\etapK, with branching fractions $\BetapKp = \RetapKp$ and
$\BetapKz = \RetapKz$.  We have searched with comparable sensitivity for
17 related decays to final states containing an $\eta$ or $\eta^\prime$
meson accompanied by a single particle or low-lying resonance.  Our
upper limits for these constrain theoretical interpretations of the
\etapK\ signal.

\end{abstract}
\newpage

{
\renewcommand{\thefootnote}{\fnsymbol{footnote}}

\begin{center}
B.~H.~Behrens,$^{1}$ W.~T.~Ford,$^{1}$
A.~Gritsan,$^{1}$ H.~Krieg,$^{1}$ J.~Roy,$^{1}$
J.~G.~Smith,$^{1}$ M.~Zhao,$^{1}$
J.~P.~Alexander,$^{2}$ R.~Baker,$^{2}$ C.~Bebek,$^{2}$
% ...
\end{center}
 
\small
\begin{center}
$^{1}${University of Colorado, Boulder, Colorado 80309-0390}\\
$^{2}${Cornell University, Ithaca, New York 14853}\\
% ...
\end{center}

\setcounter{footnote}{0}
}
\newpage

\begin{figure}[htbp]
\psfile{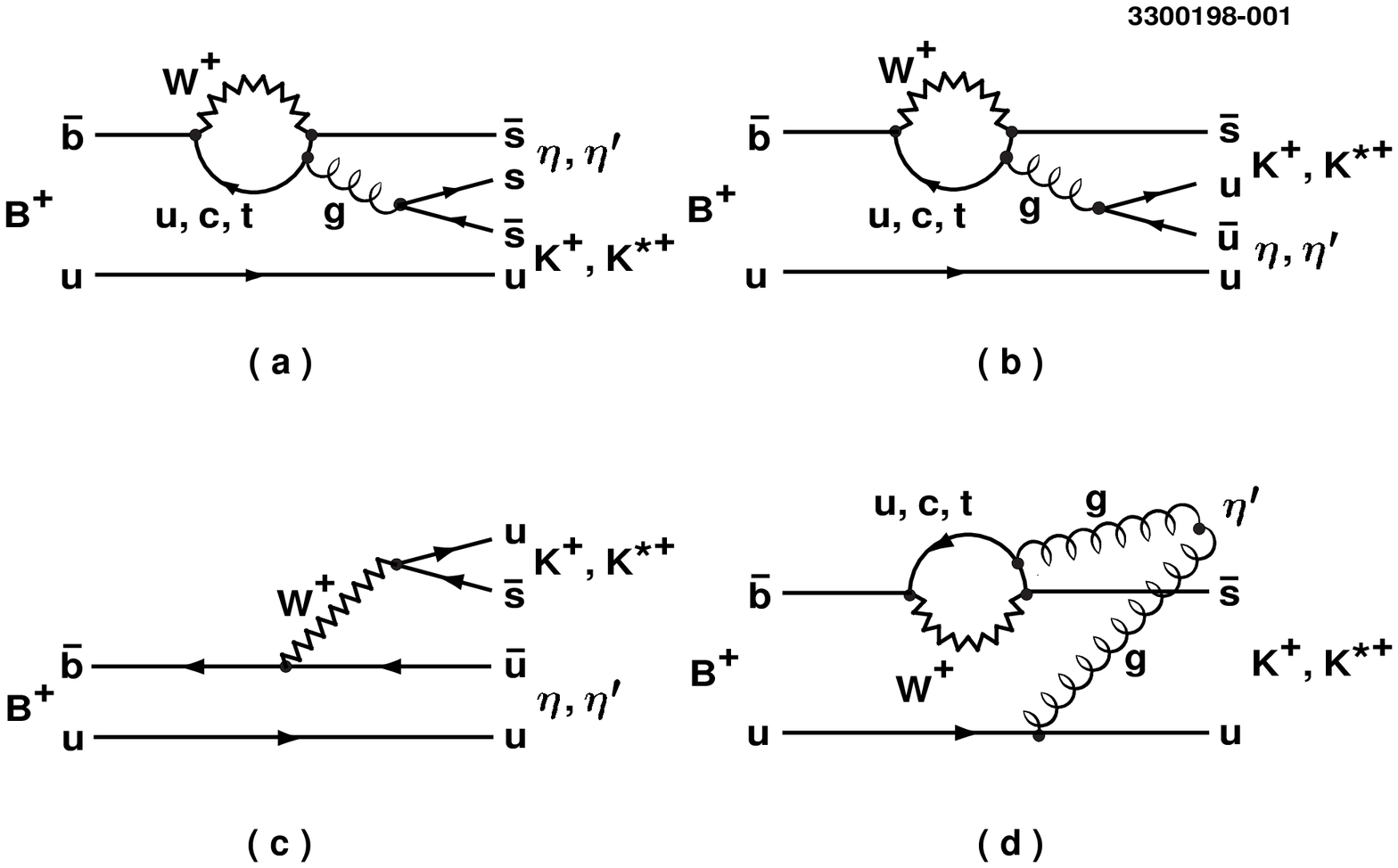}{0.9}
 \caption{\label{fig:diagrams}%
Feynman diagrams describing the representative decays
$B^+\goto\eta^{(\prime)}K^{(*)+}$:  (a, b) internal penguins; (c)
external tree; (d) flavor-singlet penguin.
 }  
\end{figure}

The dominant decay modes of $B$ mesons involve the
$\bar b\goto \bar c$ quark 
transition with coupling to a $W^+$ boson.  For many of these modes
the decay amplitude may be described by a tree diagram in which the
light quark (spectator) is bound in both the initial $B$ meson and final
charmed hadron via soft gluon exchange.  With recent improvements in
experimental sensitivity, less favored modes are becoming accessible.
These include: $b\goto u$ tree diagram transitions that are suppressed
by the 
small
Cabibbo-Kobayashi-Maskawa \cite{ckm} (CKM) matrix 
element $V_{ub}$,
such as $B\goto\pi\ell\nu$
\cite{CLEObpilnu}; effective flavor changing neutral current (FCNC)
decays $b\goto s$ described by loop diagrams, such as the ``electromagnetic
penguin'' $B\goto K^*\gamma$ \cite{CLEObkstg}; and
decays to charmless hadrons such as $B\goto K\pi$
\cite{CLEObkpi,bigrare,CLEObkpiNew}.  The 
hadronic decays 
may be classified according to contributions to the amplitude from the
several tree and penguin diagrams shown in Fig.\ \ref{fig:diagrams}
\cite{thyPred,aliGreub}.  Some of these charmless hadronic decays offer
prospects for the observation of $CP$ violation, while others
facilitate the quantitative understanding of the amplitudes that are
essential to the interpretation of future $CP$ measurements.  For
example, the decays \etaK\ and \etapK, with $B\goto K\pi$, have been
examined in this context \cite{kps,etaCP}.

In this paper we present results of experimental searches for $B$ meson
decays to two-body final states containing $\eta$ and $\eta^\prime$
mesons.  These $I=0$ mesons are mixtures of flavor-SU(3) octet and
singlet states, the latter being of particular interest because of its
allowed formation through a pure (two or more) gluon intermediate state
(Fig.\ \ref{fig:diagrams} (d)). 

The data were accumulated at the Cornell Electron-positron Storage Ring
(CESR).  The integrated luminosity was 3.11 ${\rm fb}^{-1}$ for the
reaction $e^+e^-\goto\Upsilon(4S)\goto B\bar B$ (center-of-mass energy
$E_{\rm cm}=10.58$ GeV). This luminosity corresponds to the production
of $3.3\times10^6$ charged and an approximately equal number of neutral
$B$ mesons.
In addition we recorded 1.61 ${\rm fb}^{-1}$ of data with
$E_{\rm cm}$ below the threshold for $B\bar B$ production
to measure continuum processes.

The CLEO II detector\cite{CLEOdet}\ emphasizes precision charged
particle tracking, with specific ionization ($dE/dx$) measurement, and
high resolution electromagnetic calorimetry 
based on CsI(Tl).  From the raw data we reconstruct charged pions and
kaons, photons (from $\pi^0$, $\eta$, and 
$\eta^\prime$ decays), 
and $\pi^+\pi^-$ pairs that intersect at a vertex displaced by at least
3 mm from the collision point (``vees'', from $K^0_s\goto\pi^+\pi^-$).
Candidate $B$ decay tracks
must meet specifications on the number of
drift chamber measurements, goodness of fit, and
consistency with an origin at the primary or particular secondary
vertex.  
Candidate photons must be isolated calorimeter clusters with a
photon-like spatial distribution and energy deposition exceeding 30
MeV.  We 
exclude photon pairs from extremely asymmetric $\pi^0$ or $\eta$ decays
to reject soft photon backgrounds, requiring
$\left|\cos\theta^*\right|<0.97$, where $\theta^*$ is the meson center
of mass decay angle relative to its flight direction.  We reject charged
tracks and photon pairs having momentum less than $100\ \MeVc$.

We fit photon pairs and vees kinematically to the appropriate combined
mass hypothesis to obtain the meson momentum vectors.  Resolutions on
the reconstructed masses prior to the constraint are about $5-10\
\MeVcsq$ (momentum dependent)
for $\pi^0\goto\gamma\gamma$, $12\ \MeVcsq$ for $\eta\goto\gamma\gamma$,
and $3\ \MeVcsq$ for $K^0_s\goto\pi^+\pi^-$.  Information about expected
signal distributions with the detector response comes from a detailed
GEANT based simulation of the CLEO detector \cite{GEANT}\ that
reproduces the 
resolutions and efficiencies of data in a variety of benchmark
processes.  

Since the $B$ mesons are formed nearly at rest, while the $B$ daughters
we observe are relatively light, the latter have momenta close to half
of the beam energy ($2.6\ \GeVc$).  For this reason the final states are
well separated 
from those involving heavier daughters, i.e., the dominant $b\goto c$
decays.  
The principal signatures for the selected decay modes are
consistency of the resonance decay invariant masses with
the known masses and widths of those resonances, and kinematic
consistency of the total
final state with the $B$ meson (mass and energy).  
Because
the beam energy $E_b$ is better known than the reconstructed $B$ meson
energy $E_B$, we substitute the former in the $B$ mass calculation: $\mb
\equiv \sqrt{E_b^2-{\bf p}_B^2}$, with ${\bf p}_B$ the reconstructed $B$
momentum.  We define also the variable $\Delta E\equiv E_B-E_b$.  The
measurement resolution on \mb\ is about $2.6\ \MeVcsq$, and on \DE\ it is
25-40 MeV, depending on the apportionment of the energy among charged
tracks and photons for each mode.

For vector-pseudoscalar decays of the $B$ and $\rho\gamma$ decays of the
$\eta^\prime$ we gain further discrimination from the
helicity variable \hel\ (cosine of the vector meson's rest frame
two-body decay angle relative to its flight direction), which
reflects the spin alignment in the decay. 
For modes in which one daughter is a single
charged track, or is a resonance pairing a charged track with a $\pi^0$, we
achieve statistical discrimination between kaons and pions by $dE/dx$.
With $S_K$ and $S_\pi$ defined as the
deviations from nominal energy loss for the indicated particle
hypotheses measured in standard deviations, the separation $S_K-S_\pi$
is about 1.7 at $2.6\ \GeVc$. 

The main backgrounds arise from continuum quark production
$e^+e^-\goto q\bar q$.  We discriminate against these jet-like events
with several measures of the energy flow pattern.  One is the angle
$\theta_{BB}$ between the thrust axis (axis of maximum energy projection
magnitude) of the candidate $B$ and that of the rest of the event.  For
a fake $B$ candidate selected from particles belonging to a $q\bar q$
event those particles tend to align with the rest of the event, whereas
the true $B$ decays have a thrust axis that is largely uncorrelated with the
tracks and showers from the decay of the partner $B$.  We reject events
with $\left|\cos\theta_{BB}\right|>0.9$.  In addition we use a
multivariate discriminant \xf\ incorporating the energy
deposition in nine cones concentric with the event thrust axis, and the
angles of the thrust axis and ${\bf p}_B$ with respect to the $e^+e^-$
beam direction \cite{bigrare}.  We have checked the
backgrounds from the favored $B$ decay modes by simulation and found their
contributions to the modes in this study to be negligible.

\begin{table}[htbp]
\vbox{\vspace{-1.0cm}
\caption{Measurement results.  Columns list the final states (with
secondary decay modes as subscripts), event yield from the fit,
reconstruction efficiency $\epsilon$, total efficiency with secondary
branching fractions $\calB_s$, and the resulting $B$ decay branching
fraction \calB.  }
\vspace{-2mm} 
\def\notext{ & & & & \cr}
\begin{center}
\begin{tabular}{lcrrc}
Final state&Fit events&$\epsilon$(\%)&$\epsilon\calB_s$(\%)&\calB($10^{-5})$\cr
\sgline
\etaprkpd     &$11.2^{+4.1}_{-3.4}$ &30&5.1&$6.7^{+2.5}_{-2.1}\pm0.8$\cr
\etaprkprg    &$19.6^{+6.6}_{-5.7}$ &28&8.4&$7.0^{+2.4}_{-2.1}\pm0.9$\cr
\etaprkpfv    & $2.3^{+2.2}_{-1.5}$ &17&1.7&$4.2^{+4.0}_{-2.7}\pm1.4$\cr
\etaprkzd     & $1.4^{+1.7}_{-1.0}$ &23&1.4&$3.1^{+3.7}_{-2.1}\pm0.6$\cr
\etaprkzrg    & $5.7^{+3.7}_{-2.8}$ &27&2.8&$6.2^{+4.0}_{-3.0}\pm1.2$\cr
\etaprpid     & $1.4^{+2.2}_{-1.4}$ &30&      5.2      & $<3.7$\cr
\etaprpirg    & $4.0^{+4.6}_{-3.3}$ &29&      8.8      & $<4.5$\cr
\etaprpifv    & $0.5^{+1.9}_{-0.5}$ &18&      1.8      & $<10.7$\cr
\etaprpizepp  & $0.0^{+0.5}_{-0.0}$ &25&      4.3      & $<1.8$\cr
\etaprpizrg   & $0.0^{+2.0}_{-0.0}$ &29&      8.7      & $<2.2$\cr
\etapretaprd  & $0.0^{+0.5}_{-0.0}$ &19&      0.6      &$<15.2$\cr
\etapretaprrg & $0.0^{+0.8}_{-0.0}$ &19&      1.7      & $<6.4$\cr
\etapretagg   & $0.0^{+0.5}_{-0.0}$ &26&      1.8      & $<4.6$\cr
\etapretathrp & $0.0^{+0.5}_{-0.0}$ &17&      0.7      &$<12.5$\cr
\etapretarg   & $5.6^{+4.6}_{-3.6}$ &28&      3.3      &$<13.0$\cr
\etapretargtp & $0.0^{+0.6}_{-0.0}$ &16&      1.1      & $<9.3$\cr
\etaprkstpd   & $0.0^{+1.0}_{-0.0}$ &13&      0.7      &$<18.$\cr
\etaprkstpkz  & $0.0^{+1.6}_{-0.0}$ &15&      0.6      &$<24.$\cr
\etaprkstzd   & $0.0^{+0.7}_{-0.0}$ &22&      2.5      & $<3.9$\cr
\etaprrhopd   & $0.0^{+0.7}_{-0.0}$ &12&      2.0      & $<5.7$\cr
\etaprrhozd   & $0.0^{+0.5}_{-0.0}$ &22&      3.8      & $<2.3$\cr
\etakgg       & $1.3^{+3.5}_{-1.3}$ &46&     17.9      & $<1.5$\cr
\etakthrp     & $0.0^{+2.5}_{-0.0}$ &28&      6.3      & $<3.1$\cr
\etakzgg      & $1.8^{+2.4}_{-1.6}$ &32&      4.2      & $<4.7$\cr
\etakzthrp    & $0.0^{+0.5}_{-0.0}$ &14&      1.1      & $<8.6$\cr
\etapigg      & $0.2^{+5.0}_{-0.2}$ &47&     18.2      & $<1.7$\cr
\etapithrp    & $0.0^{+1.8}_{-0.0}$ &29&      6.6      & $<2.6$\cr
\etapizgg     & $0.0^{+0.9}_{-0.0}$ &33&     13.0      & $<0.9$\cr
\etapizthrp   & $0.0^{+1.5}_{-0.0}$ &23&      5.5      & $<2.7$\cr
\etaetagg     & $1.1^{+1.7}_{-1.1}$ &34&      5.2      & $<3.0$\cr
\etaetathrp   & $0.0^{+1.3}_{-0.0}$ &24&      4.3      & $<2.9$\cr
\etaetasixp   & $0.0^{+0.5}_{-0.0}$ &16&      0.8      & $<9.8$\cr
\etakstpgg    & $0.7^{+3.6}_{-0.7}$ &25&      3.3      & $<8.8$\cr
\etakstpthrp  & $0.0^{+1.2}_{-0.0}$ &15&      1.2      & $<11.7$\cr
\etakstpggkz  & $0.0^{+1.2}_{-0.0}$ &24&      2.1      & $<5.7$\cr
\etakstpthrpkz& $0.0^{+1.0}_{-0.0}$ &14&      0.8      &$<16.0$\cr
\etakstzgg    & $5.2^{+4.0}_{-3.0}$ &32&      8.4      & $<4.6$\cr
\etakstzthrp  & $0.0^{+0.8}_{-0.0}$ &20&      3.1      & $<3.6$\cr
\etarhopgg    & $1.2^{+4.1}_{-1.2}$ &24&      9.9      & $<3.3$\cr
\etarhopthrp  & $2.5^{+4.1}_{-2.5}$ &14&      3.3      & $<11.2$\cr
\etarhozgg    & $0.2^{+4.0}_{-0.2}$ &36&     14.3      & $<1.9$\cr
\etarhozthrp  & $0.0^{+1.1}_{-0.0}$ &22&      5.1      & $<2.7$\cr
\end{tabular}
\end{center}
\label{individtab}
}
\end{table}

To extract event yields we perform unbinned maximum likelihood fits to the
data, including sidebands about the expected mass and energy peaks,
of a superposition of expected signal and background distributions:
\begin{equation}\label{eq:lfit}
 \calL(N_S,N_B) = e^{-(N_S+N_B)} \prod_{i=1}^N \left[N_{S}
               \calP_{S}(\vbeta;{\bf x}_i) + 
               N_B \calP_{B}(\vgamma;{\bf x}_i)\right].
\end{equation}
Here $\calP_S$ and $\calP_B$ are the probability distribution
functions (PDFs) for signal and continuum background, respectively.
They are functions of observables ${\bf x}_i$ for event $i$, and of
parameters \vbeta\ and \vgamma\ (discussed below).
The form of \calL\ reflects the underlying Poisson statistics obeyed by
$N_S$ and $N_B$, the (positive-definite) numbers of signal and continuum
background events, respectively, 
whose expectation values sum 
to the total number $N$ of input
events.  Observables for each event include \mb, \DE, \xf, and 
(where applicable) resonance masses and \hel.  Where two modes involve
a charged hadron (generically $h^+$) that is either  
$\pi^+$ or $K^+$ we fit both simultaneously, with
\calL\ expanded so that the signal and background
yields of both $\pi^+$ and $K^+$ are fit variables.  In this case
the PDFs depend also on
the $dE/dx$ observables $S_\pi$ and $S_K$.  The number of events $N$
for these fits ranges from $\sim 30$ to a few thousand.

The PDFs $\calP_S$ and $\calP_B$ are constructed as products of
functions of the observables ${\bf x}_i$.  The dependences of $\calP_S$ on
masses and energies are Gaussian, double Gaussian, or Breit-Wigner
functions, whose means, widths, etc.\ appear as the parameters
\vbeta\ in Eq.\ \ref{eq:lfit}.  The background PDF $\calP_B$ contains
signal-like peaking 
components in its resonance mass projections, to account for real
resonances in the background, added to smooth components for
combinatoric continuum.  The smooth components are low-order polynomials,
except that for \mb\ we use an empirical shape
\cite{argus}\ that accounts for the phase space limit at $M=E_b$.  
The dependences of both $\calP_S$ and $\calP_B$ on \xf, $S_K$, and
$S_\pi$ are bifurcated Gaussian functions.  We obtain the parameters
\vbeta\ of $\calP_S$ from separate fits to simulated signal, and \vgamma\ of
$\calP_B$ from fits to data in a sideband region of the $\DE-\mb$
plane.

Results for our 42 $B$ decay chains \cite{chgconj}\ appear
in Table \ref{individtab}.  The row label subscripts denote secondary
decays, including $\eta^\prime\goto \eta\pi^+\pi^-$ with
$\eta\goto\gamma\gamma$ ($\eta\pi\pi$), $\eta^\prime\goto \eta\pi^+\pi^-$ with
$\eta\goto\pi^+\pi^-\pi^0$ ($5\pi$), and 
$\eta\goto\pi^+\pi^-\pi^0$ ($3\pi$).  The table gives each branching
fraction quoted as 
central value with statistical followed by systematic error, or as 90\%
confidence level upper limit.  We include systematic errors from
uncertainties in the PDFs, i.e., in \vbeta\ and \vgamma, obtained from a 
Monte Carlo convolution of the likelihood function with Gaussian resolution
functions for these parameters, including their most 
important correlations.  This procedure changes the upper limit by less
than 10\% in most cases.  We also include systematic errors for
reconstruction efficiencies and selection requirements, and quote upper
limits computed with efficiencies one standard deviation below nominal.

\begin{table}[htbp]
\caption{Combined branching fraction results, with expectations from
theoretical models.}
\def\notext{ & & \cr}
\begin{center}
\begin{tabular}{lcl}
Decay mode &    \calB($10^{-5})$      & Theory \calB($10^{-5}$)\cr
\sgline
$B^+\goto\etaprkp$    & $6.5^{+1.5}_{-1.4}\pm0.9$&$0.7-4.1$\cite{kps,chau,du}\cr
$B^0\goto\etaprkz$    & $4.7^{+2.7}_{-2.0}\pm0.9$&$0.9-3.3$\cite{chau,du}\cr
$B^+\goto\etaprpi$    &          $<3.1$          &$0.8-3.5$\cite{kps,chau,du}\cr
$B^0\goto\etaprpiz$   &          $<1.1$          &$0.4-1.4$\cite{chau,du}\cr
$B^0\goto\etapretapr$ &          $<4.7$          &$0.1-2.8$\cite{chau,du}\cr
$B^0\goto\etapreta$   &          $<2.7$          &$0.4-4.4$\cite{chau,du}\cr
$B^+\goto\etaprkstp$  &          $<13.$          &$0.1-0.9$\cite{kps,chau,du}\cr
$B^0\goto\etaprkstz$  &          $<3.9$          &$0.8-1.7$\cite{chau,du}\cr
$B^+\goto\etaprrhop$  &          $<4.7$          &$0.8-5.7$\cite{kps,chau,du}\cr
$B^0\goto\etaprrhoz$  &          $<2.3$          &$0.2-1.2$\cite{chau,du}\cr
$B^+\goto\etak$       &          $<1.4$          &$0.1-0.5$\cite{kps,chau,du}\cr
$B^0\goto\etakz$      &          $<3.3$          &$0.1-0.2$\cite{chau,dean,du}\cr
$B^+\goto\etapi$      &          $<1.5$          &$0.2-0.8$\cite{kps,chau,dean,du}\cr
$B^0\goto\etapiz$     &          $<0.8$          &$0.2-0.4$\cite{chau,du}\cr
$B^0\goto\etaeta$     &          $<1.8$          &$0.1-1.4$\cite{chau,dean,du}\cr
$B^+\goto\etakstp$    &          $<3.0$          &$0.1-1.3$\cite{kps,chau,du}\cr
$B^0\goto\etakstz$    &          $<3.0$          &$0.1-0.5$\cite{chau,dean,du}\cr
$B^+\goto\etarhop$    &          $<3.2$          &$0.7-4.4$\cite{kps,chau,dean,du}\cr
$B^0\goto\etarhoz$    &          $<1.3$          &$0.1-0.8$\cite{chau,dean,du}\cr
\end{tabular}
\end{center}
\label{combtab}
\end{table}

Where we have measured a given $B$ decay mode in more than one secondary
decay channel we combine the samples by adding the $\chi^2=-2\ln\calL$ 
functions of branching fraction and extracting a value with errors
or limit from the combined distribution.  The limit is the value of
\calB\ below which 90\% of the integral of \calL\ lies.  The
results are summarized in Table \ref{combtab}, together with previously
published theoretical calculations \cite{kps,chau,dean,du}.

We have analyzed each of the decays also without use of the likelihood
fit, employing more restrictive cuts in each of the variables to isolate
the signals.  The results are consistent with those quoted in the
tables, but with larger errors (less restrictive limits) in most cases.

\begin{figure}[htbp]
\psfile{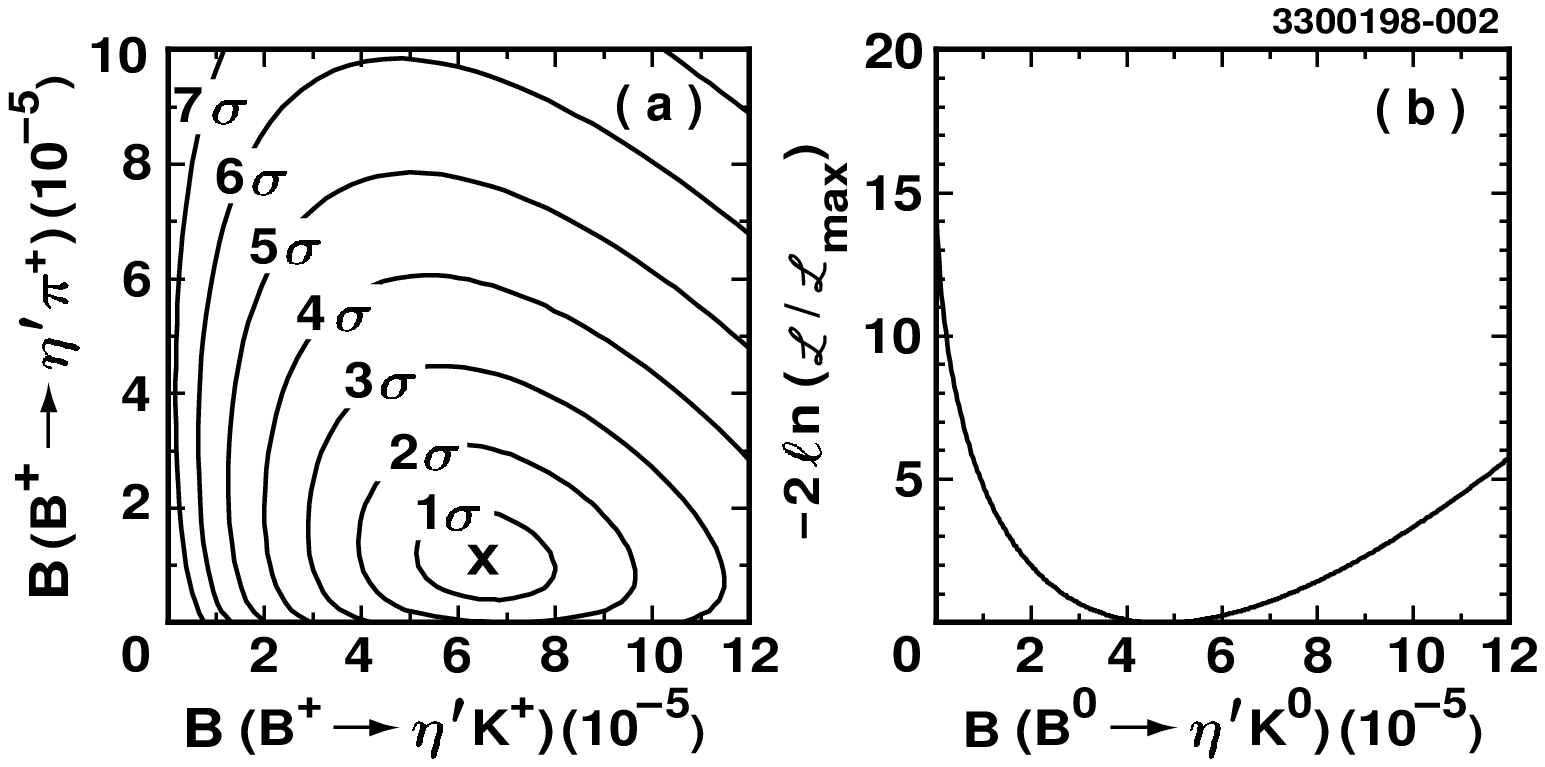}{1.0}
 \caption{\label{fig:etapCont}%
(a) Likelihood function contours for $B^+\goto\eta^\prime h^+$; (b)
The function $-2\ln{\calL/\calL_{\rm max}}=\chi^2-\chi^2_{\rm min}$ for
\etapKz. 
 }  
\end{figure}

\begin{figure}
\psfile[110 465 535 705]{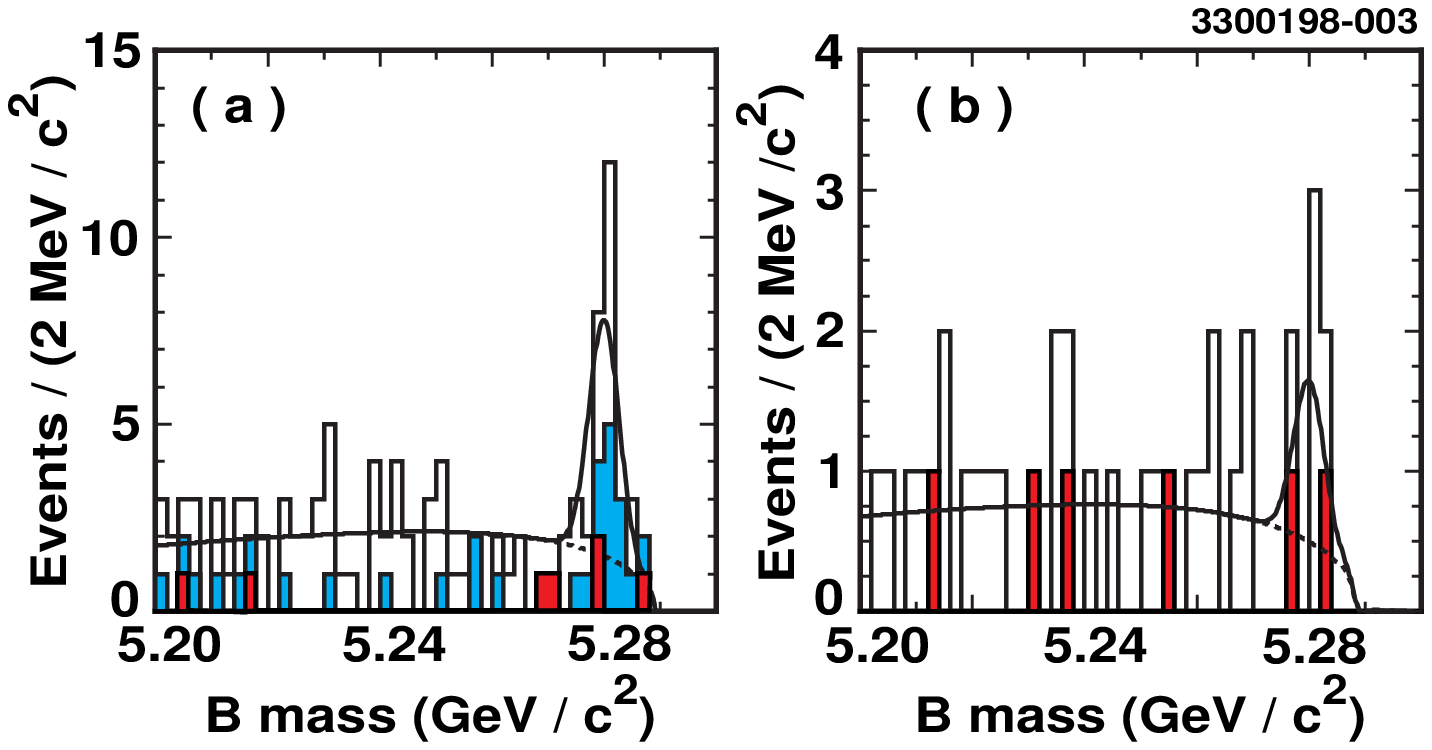}{1.0}
  \caption{\label{fig:etaphMbDe}%
Projections onto the variable \mb.  Overlaid on each plot as smooth
curves are the best fit functions (solid) and background components
(dashed), calculated with the variables not shown restricted to the
neighborhood 
of expected signal.  The histograms show (a) $B^+\goto\eta^\prime
h^+$ with $\eta^\prime\goto \eta \pi\pi\ (\eta\goto3\pi$, dark shaded), 
$\eta^\prime\goto \eta \pi\pi\ (\eta\goto\gamma\gamma$, light shaded),
and $\eta^\prime\goto \rho\gamma$ (open); (b) \etapKz\ with
$\eta^\prime\goto \eta \pi\pi$ (shaded) and $\eta^\prime\goto
\rho\gamma$ (open). 
 }
\end{figure}

We find positive signals in both charge states of $B\goto\eta^\prime K$:
$\BetapKp = \RetapKp$ and $\BetapKz = \RetapKz$.
(The first error quoted is statistical, the second systematic.)  
The significance, defined as the number of standard deviations
corresponding
to the probability for a fluctuation from zero to our observed
yield, is 7.5 for \etapKp and 3.8 for $B^0\goto\eta^\prime K^0$.  The
likelihood functions from the fits for $B\goto \eta^\prime h^+$ and
$B^0\goto\eta^\prime K^0$ are shown in Fig.\ \ref{fig:etapCont}.  For
these modes we show also in Fig.\ \ref{fig:etaphMbDe}\ the projections
of event distributions onto the \mb\
axis.  Clear peaks at the $B$ meson mass are evident.

The observed branching fractions for \etapK, in combination with the
upper limits for the other modes in Table \ref{combtab} and with recent
measurements of $B\goto K\pi$ and $B\goto\pi\pi$ \cite{CLEObkpiNew},
provide important constraints on the theoretical picture for these
charmless hadronic decays.  A large
ratio of \etapK\ to \etaK, consistent with our measurements, was
predicted \cite{lipkin} in terms of
interference of the two penguin diagrams in Fig.\
\ref{fig:diagrams}(a) and (b), constructive for \etapK\
and destructive for \etaK.  
The effective Hamiltonian calculations \cite{aliGreub}\ contain
uncertainties in form factors \cite{datta,kagan}, light quark masses
\cite{kagan}, the QCD scale, and the
effective number of colors.  They generally employ spectator and factorization
\cite{BSW} approximations.  The unexpectedly large branching fraction for
\etapK\ has led to a reevaluation of some of the older calculations.
Recent suggestions include contributions from the QCD gluon anomaly
or other flavor singlet processes (Fig.\ 
\ref{fig:diagrams} (d)) in constructive interference with the penguins
\cite{etaCP,soni,hairpin,cheng,zhitnitsky}.  Prospects are good for
resolution of some of these issues as new data become
available.

We thank A. Ali, T. DeGrand, P. Lepage, H. Lipkin, A. Kagan, and A. Soni 
for useful discussions.
We gratefully acknowledge the effort of the CESR staff in providing us with
excellent luminosity and running conditions.
This work was supported by 
the National Science Foundation,
the U.S. Department of Energy,
the Heisenberg Foundation,  
Research Corporation,
the Natural Sciences and Engineering Research Council of Canada, 
the A.P. Sloan Foundation, 
and the Swiss National Science Foundation.


\begin{thebibliography}{99}
\bibitem{chgconj} 
Inclusion of the charge conjugate states is implied
in the row labels and throughout this paper.
\end{thebibliography}
\end{document}